\pdfoutput=1 
\documentclass{appolb}

\usepackage{amsmath,amssymb,bbm,color}
\usepackage{epsfig}
\usepackage{graphics}
\usepackage{euscript}
\usepackage{pslatex}
\usepackage{hyperref}
\usepackage{xcolor}
\usepackage{fancybox}
\usepackage{enumerate}
\usepackage{subfig}

\newcommand{\Ecal}{{\cal E}}
\newcommand{\bk}{{\bf{k}}}

\newcommand{\tbet}{{\tilde{\beta}}}

\begin{document}


\title{
\vspace{-4.0cm}
\begin{flushright}
		{\small {\bf IFJPAN-IV-2012-6}}   \\
		{\small {\bf SMU-HEP-12-14}}  \\
\end{flushright}
\vspace{0.5cm}
NLO corrections to hard process in QCD shower -- proof of concept
\thanks{
This work is partly supported by 
the Polish Ministry of Science and Higher Education grant No.\ 1289/B/H03/2009/37,
 the Polish National Science Centre grant DEC-2011/03/B/ST2/02632,
 the NCBiR grant LIDER/02/35/L-2/10/NCBiR/2011,
  the Research Executive Agency (REA) of the European Union 
  Grant PITN-GA-2010-264564 (LHCPhenoNet),
the U.S.\ Department of Energy
under grant DE-FG02-04ER41299 and the Lightner-Sams Foundation.
}}

\author{S.\ Jadach$^a$, M.\ Je\.zabek$^a$, A.~Kusina$^b$, W.\ P\l{}aczek$^c$, M.\ Skrzypek$^a$
\address{$^a$ Institute of Nuclear Physics, Polish Academy of Sciences,\\
              ul.\ Radzikowskiego 152, 31-342 Cracow, Poland}
\address{$^b$ Southern Methodist University, Dallas, TX 75275, USA}
\address{$^c$ Marian Smoluchowski Institute of Physics, Jagiellonian University,\\
              ul. Reymonta 4, 30-059 Cracow, Poland.}
}


\maketitle

\begin{abstract}
The concept of new methodology of adding 
QCD NLO corrections in the initial state Monte Carlo parton shower
(hard process part) is tested numerically using, as an example,
the process of the heavy boson production
at hadron--hadron colliders such as LHC.
In spite of the use of a simplified model of the process,
all presented numerical results prove
convincingly that the basic concept of the new methodology
works correctly in practice, that is
in the numerical environment of the Monte Carlo parton shower
event generator.
The differences with the other well established methods,
like MC@NLO and POWHEG, are briefly discussed
and future refinements of the implementation of the new method
are also outlined.
\end{abstract}


\vspace{2mm}
\begin{center}
\em Acta Physica Polonica in press
\end{center}

\section{Introduction}

Successful operation of the Large Hadron Collider (LHC) at CERN
is resulting in rich harvest of experimental data.
Even more data at higher energy and with higher statistics
will be available over the next two decades from the LHC experiments.
One of the challenges in the proper understanding and interpretation
of these data,
possibly leading to discovery of new phenomena,
will be perfect mastering of the ``trivial'' effects due to multiple
emission of soft and collinear gluons and quarks.
Perturbative Quantum Chromodynamics
(pQCD)~\cite{GWP,Gross:1974cs,Georgi:1951sr},
together with the clever modelling of low energy
nonperturbative effects, will be the basic and indispensable tool
for disentangling the Standard Model physics component in the data.

\section{Overview of the method}

Most of the methodology used in this work is described 
in ref.~\cite{Jadach:2011cr}.
In the following, additional
details relevant for Monte Carlo (MC) implementation and tests are described.
Let us start with a description of the initial condition of 
the forward evolution (necessary in the MC implementation)
which was omitted in ref.~\cite{Jadach:2011cr}.
For the proper understanding of this implementation it is necessary
to recall some basic facts about the use of maximum rapidity of emitted
partons (angular ordering) as the evolution time variable.

While the energy of the emitted gluon is a natural variable to handle
infrared singularities,
the angular variable is best suited for controlling collinear singularities.
The logarithm of the angle of the emitted gluon (rapidity) with respect to
the {\em emitter parton} emerging from the initial hadron,
is a natural ``master variable'' for modelling collinear singularities.
The angular variable is also well suited
for modeling the structure of the non-abelian soft
limit (colour coherence)~\cite{Ermolaev:1981cm,khoze-book,Slawinska:2009gn}.

Conversely,
a hard process in deep inelastic lepton-hadron scattering (DIS)
or Drell-Yan (DY) process acting as a ``probe'',
either backscattering (in the Breit frame for DIS)
or absorbing (into a heavy boson in DY) the emitter parton,
in a well defined rest frame of the hard process (RFHP),
has its own energy scale used also as a master variable in collinear
factorization and renormalization group equations.
More precisely,
for the hard process with the center-of-mass energy $Q$
($\sqrt{\hat{s}}$ in the DY process),
the parton entering hard process has energy equal $Q/2$.
On the other hand, in the same RFHP,
the initial hadron energy is $E_h$ and
the Bjorken variable is the ratio $x=Q/(2E_h)$
($x$ is invariant with respect to boosts along the emitter direction).
The luminosity distribution of this parton $D(Q,x)$ is
commonly referred to as parton distribution function, PDF in short.
It is weakly dependent on $Q$ and is measured experimentally
at each value of $Q$ separately,
that is at a given value of $Q$
varying the energy $E_h=Q/(2x)$ of the initial hadron seen in RFHP.

The important practical question for Monte Carlo modelling of the emission
of the collinear gluons is: how to relate
the variable $\hat{t}=\ln(Q/\Lambda)$
governing the ``evolution'' of the PDF
in the traditional DGLAP schemes%
\footnote{Formulated typically in terms of exponentiation of the collinear
 singularities or using the renormalization group equations, or both.
}
such as $\overline{MS}$ scheme,
and the rapidity variable of emitted gluons?

In the following, to answer the above question
we shall consider for simplicity the LO case with non-running $\alpha_S$,
and for pure gluonstrahlung (non-singlet QED-like component of PDF)
from a {\em single emitter}.%
\footnote{One hemisphere in DY, 
          or initial state cascade/ladder in DIS process.}
For the MC purpose, we define the evolution variable $t$
as a hypervelocity of the Lorentz boost from the initial beam hadron
rest frame to RFHP, $t=\Xi$
(or equivalently, the hypervelocity of the beam hadron in the RFHP).
For each emitted gluon we define the rapidity $\xi_i$ in the rest frame
of the initial hadron, or $\hat{\xi}_i$ in RFHP.
Next, we require in the context of pQCD description
of the gluonstrahlung that $\hat{\xi}_i<0$ in the RFHP
(in the initial hadron frame $\xi_i<\Xi$).
In other words, $t=\Xi$
is a limiting rapidity for emitted gluons (or $\hat\xi=0$).
Of course,  $e^t=\frac{\sqrt{s}}{m_h}= \frac{Q}{x m_h}$,
where $m_h$ is hadron mass and $s=4E_h^2$.

The role of perturbative QCD is to relate $D(Q,x)$ measured
in two experiments A and B with probes at the scales $Q_A$ and $Q_B$,
provided that $Q_A>Q_B >> m_h$. 
Two PDFs, $D(Q_A,x)$ and $D(Q_B,x)$, will differ
because of the gluon emissions
located in the additional phase space within the $(\Xi_A,\Xi_B)$
rapidity (angular) interval.
Also, experiments A and B will use different RFHPs,
connected by the Lorentz boost of the hypervelocity
$\Delta t =\Xi_A-\Xi_B =\ln\frac{Q_A x_B}{Q_B x_A}$.

What is now the difference between
the more traditional choice of the evolution time variable
$\hat{t}=\ln\frac{Q}{\Lambda}$ of DGLAP 
and our preferred definition 
$t=\Xi=\ln\frac{E_h}{m_h}\big|_{\rm RFHP}=\ln\frac{Q}{x m_h}$
(maximum rapidity of the emitted gluons)?
When comparing two experiments with hard probes
at the scales $Q_A$ and $Q_B$,
$\Delta \hat{t}= \ln(Q_A/Q_B)$, 
while more phase-space conscious 
$\Delta t= \Delta \hat{t} -\ln(x_A/x_B)$.
The ``offset'' $\ln(x_A/x_B)$
is formally of the NLO class%
\footnote{
 It induces extra ${\cal O}(\alpha_S)$ term in the evolution kernel.
}
and can be neglected within the LO approximation,%
\footnote{We have to remember to take it into account
at the NLO level, when defining NLO evolution kernel.}
hence both choices are equally good at LO level.
However, the use of (angular) $t$ assures the completeness of the phase space
of the emitted gluon, no gaps (nor ``dead zones''),
so it is the preferred choice in the MC modelling,
aiming at the NLO level evolution in the next steps.
Additionally,
the parallel use of $\hat{t}=t-\ln\frac{\Lambda}{x m_h}$ 
is quite useful and essential for other purposes,
like introduction of the running $\alpha_S$, etc.

In particular, $\hat{t}$ is more natural for defining the initial point
of the forward evolution (the stopping rule in the backward evolution).
In order to assure the validity of pQCD it is
required that the energy scale of the probe $q_0>>\Lambda,m_h$
is reasonably above
the non-perturbative scales, like $\Lambda\simeq m_h\simeq1$GeV,
at the above initial point.
This leads to the initial forward evolution point at
$\hat{t}_0 \simeq \ln(q_0/\Lambda)$ and $t_0 \simeq \ln(q_0/m_h) -\ln x_0$,
as implemented in the following MC.
In other words,
gluons with rapidities below $t_0$ are regarded as ``unresolved'', 
i.e. $t_0=\xi_0$ is a maximum rapidity for all unresolved gluons.

It should be noted that the above discussion is quite standard
in the context of any Monte Carlo parton shower using angular ordering.
This line of the MC parton shower inspired by CCFM model~\cite{CCFM},
see also refs.~\cite{Marchesini:1995wr,Catani:1990rr,Marchesini:1988cf},
is presently developed by CASCADE MC authors~\cite{Jung:2010si}.
In particular,
when using (maximum) rapidity $t$ as the evolution time variable
in the time ordered exponential of
the QCD parton distributions the complete multigluon
phase space is covered (with no gaps, ``dead zones''),
while the straightforward use of the ordering in the $\hat{t}$ variable
in the MC would result in gaps between emitted real hard gluons,
see also brief discussion of the corresponding
kinematics in ref.~\cite{Jadach:2007qa}.

\subsection{Single LO ladder -- basic building block in the MC}
Let us define multigluon distribution in the single initial state
ladder taken in the LO approximation, which is a building block
in our parton shower MC implementation,
as an integrand in the following 
``exclusive/unintegrated PDF'':
\begin{equation}
\label{eq:LOMC}
\begin{split}
& 
D(t,x)
=\int dx_0\; dZ\; 
 \delta_{x=x_0 Z}\;
d_0(\hat{t}_0,x_0)\; G(t, \hat{t}_0- \ln x_0 | Z),
\\&
G(t, t_0 | Z)=
 e^{-S_F} \sum_{n=0}^\infty
\bigg(
\prod_{i=1}^n
 \int d^3\Ecal(\bar{k}_i)\;
 \theta_{\xi_i>\xi_{i-1}}
 \frac{2C_F\alpha_s}{\pi^2} \bar{P}(z_i)
\bigg)
\\&
\qquad\qquad\quad\times
\theta_{t>\xi_n}
\delta_{z=\prod_{j=1}^n z_j},
\end{split}
\end{equation}
where $\bar{P}(z)=\frac{1}{2}(1+z^2)$, $\hat{t}_0=\ln(q_0/\Lambda)$.
The ``eikonal''  phase space integration element is defined as
in ref.~\cite{Jadach:2011cr}%
\footnote{
  A single ladder (parton shower from single emitter) is defined
  in the ``tanget space'' of momenta $\bar{k}$, see ref.~\cite{Jadach:2011cr}.
}
\[
d^3\Ecal(k)=\frac{d^3 k}{2k^0}\;\frac{1}{\bk^2}
            =\pi \frac{d\phi}{2\pi} \frac{d k^+}{k^+} d \xi
\]
and $k^\pm = k^0\pm k^3$.
In the above we use
rapidity $\xi =\frac{1}{2}\ln\frac{k^-}{k^+}\big|_{\rm Rh}$ 
defined in the beam hadron rest frame Rh,
while $\eta=\frac{1}{2}\ln\frac{k^+}{k^-}\big|_{\rm RFHP}$
of ref.~\cite{Jadach:2011cr}
was defined in laboratory frame.%
\footnote{
  Or, alternatively, in the overall center of the mass system.}
They are simply related by $\xi=\ln\frac{\sqrt{s}}{m_h}-\eta$.
Rapidity  ordering is now 
$t=\xi_{\max}>\xi_n>\dots>\xi_i>\xi_{i-1}>\dots>\xi_0=t_0$,
where $t_0=\xi_0=\ln(q_0/m_h)-\ln x_0$.
The direction of the $z$ axis in the RFHP is 
traditionally pointing out towards the hadron momentum.
A lightcone variable of the emitted gluon is defined in the usual way
as $\alpha_i= \frac{2k_i^+}{\sqrt{s}}$,
of the emitter (after $i$ emissions) as
$x_i=x_0-\sum_{j=0}^{i}\; \alpha_j$,
and finally $z_i=x_i/x_{i-1}$.
The Sudakov formfactor $S_F$
is determined by the ``unitarity'' condition
\begin{equation}
\label{eq:Gxi}
\int_0^1 dZ\; G(t, t_0 | Z)=1
\end{equation}
and we omit its explicit definition, which involves
the usual cutoff $1-z_i<\epsilon$ regularizing the IR singularity
$\frac{d\alpha_i}{\alpha_i}= \frac{dz_i}{1-z_i}$.
The above feature is instrumental in the Markovian MC implementation,
which provides $D(t,x)$
for any value of $t>t_0$.

The initial distribution 
$d_0(q_0,x_0)$ can be related to experiment,
to previous steps in the MC ladder,
or to PDF in the standard $\overline{MS}$ system.
Its precise definition is not essential for the following tests
of implementation of the NLO corrections to the hard process,
hence we will define it only numerically.
We only notice that due to eq.~(\ref{eq:Gxi}) the
baryon number conservation sum rule
\[
 \int_0^1 dx\; D(t,x) = \int_0^1 dx_0\; d_0(t_0,x_0)
\]
is preserved.

Finally, note the use in eq.~(\ref{eq:LOMC})
of the rescaled four-momenta $\bar{k}^\mu$
within the ``tangent space'', as defined in ref.~\cite{Jadach:2011cr}.
The mapping $k^\mu \to \bar{k}^\mu$ can be defined%
\footnote{This mapping preserves the rapidity variable.}
once the ladders are connected together with the hard process,
back in the common standard Lorentz invariant phase space,
see ref.~\cite{Jadach:2011cr} and the following sections.

\subsection{Two-ladder LO multiparton distributions}

As a necessary introductory step to correcting the hard process to NLO level,
let us start with defining and testing our simplified
MC parton shower implementing the DY process with
two ladders and the hard process, all three in the LO approximation:%
   \footnote{In this work we adopted notation of ref.~\cite{Jadach:2011cr},
   in particular $d\tau_2(P;q_1,q_2) =
   \delta^{(4)}(P-q_1-q_2)\frac{d^3q_1}{2q_1^0}\frac{d^3q_2}{2q_2^0}$.}
\begin{equation}
\label{eq:LOMCFBmaster}
\begin{split}
&\sigma_0=
\int d x_{0F} d x_{0B}\;\;
  d_0(\hat{t}_0,x_{0F}) d_0(\hat{t}_0,x_{0B}) 
 \sum_{n_1=0}^\infty\;
 \sum_{n_2=0}^\infty
 \int dx _F\; dx_B\;
\\&~~~~~~~~~\times
e^{-S_{_F}}
\int_{\Xi<\eta_{n_1}}
\bigg(
\prod_{i=1}^{n_1}
 d^3\Ecal(\bar{k}_i)
 \theta_{\eta_i<\eta_{i-1}}
 \frac{2C_F\alpha_s}{\pi^2} \bar{P}(z_{Fi})
\bigg)
\delta_{x_F = x_{0F}\prod_{i=1}^{n_1} z_{Fi}}
\\&~~~~~~~~~\times
e^{-S_{_B}}
\int_{\Xi>\eta_{n_2}}
\bigg(
\prod_{j=1}^{n_2}
 d^3\Ecal(\bar{k}_j)
 \theta_{\eta_j>\eta_{j-1}}
 \frac{2C_F\alpha_s}{\pi^2} \bar{P}(z_{Bj})
\bigg)
\delta_{x_B = x_{0B}\prod_{j=1}^{n_2} z_{Bj}}
\\&~~~~~~~~~\times
 d\tau_2(P-\sum_{j=1}^{n_1+n_2} k_j ;q_1,q_2)\;
\frac{d\sigma_B}{d\Omega}(sx_Fx_B,\hat\theta)\;
W^{NLO}_{MC}.
\end{split}
\end{equation}
In the LO approximation we set $W^{NLO}_{MC}=1$.
This weight will be defined/restored in the next section.
In the above we use rapidity variable $\eta$,
defined in the overall center of the mass system (CMS).
Rapidity $\xi$ of eq.~(\ref{eq:LOMC}) is translated into $\eta$,
differently in the forward part (F) of the phase space
$\eta_{0F}>\eta_i>\Xi$
where we define $\xi_i=\ln\frac{\sqrt{s}}{m_h}-\eta_i$,
and in the backward (B) part
$\Xi>\eta_i>\eta_{0B}$
where $\xi_i=-\ln\frac{\sqrt{s}}{m_h}+\eta_i$ should be used.
The boundary between the two hemispheres $\Xi$ is for the moment
set to be at $\Xi=0$, but in a more sophisticated versions
of the MC will be correlated with the position of the produced
heavy boson (LO), or heavy boson and the hardest gluon (LO+NLO).%
\footnote{
  Gluon phase space is always fully covered (no gaps).}
For the initial condition in the evolution we define
$\eta_{0F}= \ln(q_0/m_h)-\ln(x_{0F})$
and
$\eta_{0B}=-\ln(q_0/m_h)+\ln(x_{0B})$.
However, for the sake of simplicity
we will set $\ln(q_0/m_h)=0$ in the following.

Phase space integration of eq.~(\ref{eq:LOMCFBmaster}) for $W^{NLO}_{MC}=1$
and using eq.~(\ref{eq:LOMC}), provides us with
the classical factorization formula
\begin{equation}
\label{eq:LOfactDY2LO}
\sigma_0 =
\int_0^1 dx_F\;dx_B\;
D_F(t, x_F)\; D_B(t, x_B)\;
\sigma_B(sx_Fx_B).
\end{equation}
In testing numerically the above formula,
the convolutions
$D_F(t, x_F)=(d_0\otimes G_F) (t, x_F)$ and 
$D_B(t, x_B)=(d_0\otimes G_B) (t, x_B)$ are
obtained from separate simple Markovian LO Monte Carlo exercises.
As was stressed in ref.~\cite{Jadach:2011cr}, the above LO formula
represents our LO MC without any approximations
and can be tested with arbitrary numerical precision.
Such a precise numerical test is demonstrated in the next section.

In eq.~(\ref{eq:LOMCFBmaster}) distributions are expressed
(similarly as in eq.~(\ref{eq:LOMC}))
in terms of the $\bar{k}^\mu$ four-momenta in the tangent space.
The mapping $\bar{k}^\mu \to k^\mu$ is understood to be exactly the same
as defined in ref.~\cite{Jadach:2011cr}, that is done simultaneously
for both hemispheres, using ordering in the variable $|\eta_i -\Xi|$.
The details of this mapping do not influence the validity 
of eq.~(\ref{eq:LOfactDY2LO}).

\subsection{Two LO ladders and NLO-corrected DY hard process}

Introduction of the NLO corrections to the hard process is done
using a single ``monolithic'' weight $W^{NLO}_{MC}$
on top of the LO distributions of eq.~(\ref{eq:LOMCFBmaster}).
In the following numerical exercises we will implement
$W^{NLO}_{MC}$ defined exactly as in ref.~\cite{Jadach:2011cr}.
Let us recall this definition in a slightly more compact notation,
for the sake of completeness:
\begin{equation}
\label{eq:NLODYMCwt}
\begin{split}
&W^{NLO}_{MC}=
1+\Delta_{S+V}
+\sum_{j\in F} 
 \frac{\tbet_1(q_1,q_2,\bar{k}_j)}%
      {\bar{P}(z_{Fj})\;d\sigma_B(\hat{s},\hat\theta)/d\Omega}
+\sum_{j\in B} 
 \frac{\tbet_1(q_1,q_2,\bar{k}_j)}%
      {\bar{P}(z_{Bj})\;d\sigma_B(\hat{s},\hat\theta)/d\Omega},
\end{split}
\end{equation}
with the NLO soft+virtual correction
$
\Delta_{V+S}
=\frac{C_F \alpha_s}{\pi}\; \left( \frac{2}{3}\pi^2 -\frac{5}{4} \right)
$
and the real correction part:
\begin{equation}
\label{eq:DYbeta1FB}
\begin{split}
&\tbet_1(q_1,q_2,k)=
\Big[
  \frac{(1-\beta)^2}{2}
  \frac{d\sigma_{B}}{d\Omega_q}(\hat{s},\theta_{F})
 +\frac{(1-\alpha)^2}{2}
  \frac{d\sigma_{B}}{d\Omega_q}(\hat{s},\theta_{B})
\Big]
\\&~~~~~~~~~~~~~~~~~~~~
-\theta_{\alpha>\beta}
 \frac{1+(1-\alpha-\beta)^2}{2}
 \frac{d\sigma_{B}}{d\Omega_q}(\hat{s},\hat\theta)
-\theta_{\alpha<\beta}
 \frac{1+(1-\alpha-\beta)^2}{2}
 \frac{d\sigma_{B}}{d\Omega_q}(\hat{s},\hat\theta).
\end{split}
\end{equation}
The above represents the exact ME
of the quark-antiquark annihilation into a heavy vector boson process
with additional single real gluon emission%
\footnote{
 We employ here the particular compact
 representation of ref.~\cite{Berends:1980jk}
 of this ME as a combination of the Born differential sections
 with the redefined scattering angle $\theta$.
}
and subtraction of the LO component already included in the LO MC.
The angle $\theta$ in the subtraction (LO) part of the Born distribution
is typically defined in the rest frame of the heavy boson,
where $\vec{q}_1+\vec{q}_2=0$,
as an angle between the decay lepton momentum $\vec{q}_1$
and the difference of momenta of the incoming quark and antiquark
$\hat\theta=\angle(\vec{q}_1,\vec{p}_{0F}-\vec{p}_{0B})$,%
\footnote{
   Other similar choices of the angle in the Born distribution
   are also perfectly valid within the LO MC.
}
while two angles in the NLO exact ME are defined precisely as
$\hat\theta_F=\angle(\vec{q}_1,-\vec{p}_{0B})$ and
$\hat\theta_B=\angle(\vec{q}_1, \vec{p}_{0F})$.
(The implementation of NLO corrections in POWHEG scheme
in ref.~\cite{Alioli:2011nr} uses the same form of the exact ME.)
Note that in the above we only need directions of the vectors
$\vec{p}_{0F}$ and $\vec{p}_{0B}$, which are the same as of the
hadron beams.
The variable $\hat{s}=s x_F x_B = (q_1+q_2)^2$
is the effective mass squared of the heavy vector boson.
Finally, we specify the lightcone variables $\alpha_j$ and $\beta_j$
of the emitted gluon for $j$ in the F and B parts of the phase space:
\[
\begin{split}
&\alpha_j=1-z_{Fj},\quad
\beta_j=  \alpha_j\; e^{2(\eta_j-\Xi)},\quad ~~
{\rm for}~~~ j\in F,
\\&
\beta_j=1-z_{Bj},\quad
\alpha_j=\beta_j\;   e^{-2(\eta_j-\Xi)},\quad
{\rm for}~~~ j\in B.
\end{split}
\]
The above relations are explained in ref.~\cite{Jadach:2011cr}
as resulting directly from the kinematical projection operators extracting
the LO part from the exact matrix element.
Note that variables $\alpha_j$ and $\beta_j$ in the above relations
are defined in terms of $\bar{k}^\mu_j$,
which do not obey the overall 4-momentum conservation.
The transformation $\bar{k}^\mu_j \to k^\mu_j$ and its inverse 
(where $k^\mu_j$ do obey 4-momentum conservation)
are defined explicitly in the above work.
Slightly improved (LO level) kinematical mapping, better suited for the NLO-corrected
hard process will be proposed at the end of section 4.

The exact phase space integration of eq.~(\ref{eq:LOMCFBmaster}) 
including $W^{NLO}_{MC}$ of eq.~(\ref{eq:NLODYMCwt})
is again possible, see ref.~\cite{Jadach:2011cr} for details,
providing a compact expression for the
total cross section:
\begin{equation}
\label{eq:DYanxch}
\begin{split}
\sigma_1 &=
\int\limits_0^1 dx_F\;dx_B\; dz\;
D_F(t, x_F)\;D_B(t, x_B)\;
\sigma_B(szx_Fx_B)
\big\{
\delta_{z=1}(1+\Delta_{S+V})
+C_{2r}(z)
\big\},
\end{split}
\end{equation}
where
$C_{2r}(z) 
=\frac{2C_F \alpha_s}{\pi}\; \left[ -\frac{1}{2}(1-z) \right]
$
was derived in ref.~\cite{Jadach:2011cr}.

\section{Numerical results}

In the following we shall first check that
the simple formula of eq.~(\ref{eq:LOfactDY2LO}) 
with two collinear PDFs
agrees numerically with the parton shower MC
of eq.~(\ref{eq:LOMCFBmaster})
with the LO hard process ($W^{NLO}_{MC}=1$).
Once the above ``LO benchmark calibration'' is successful,
we shall check numerically whether the NLO formula
of eq.~(\ref{eq:DYanxch})
agrees with the MC integration of eq.~(\ref{eq:LOMCFBmaster}),
switching on the
NLO correcting weight $W^{NLO}_{MC}$ of eq.~(\ref{eq:NLODYMCwt}).
In both MC exercises
we expect deviations only up to statistical MC error, 
or other imperfections of the numerical implementations.

\subsection{LO benchmark}

\begin{figure}
  \centering
  {\includegraphics[width=0.75\textwidth,height=85mm]{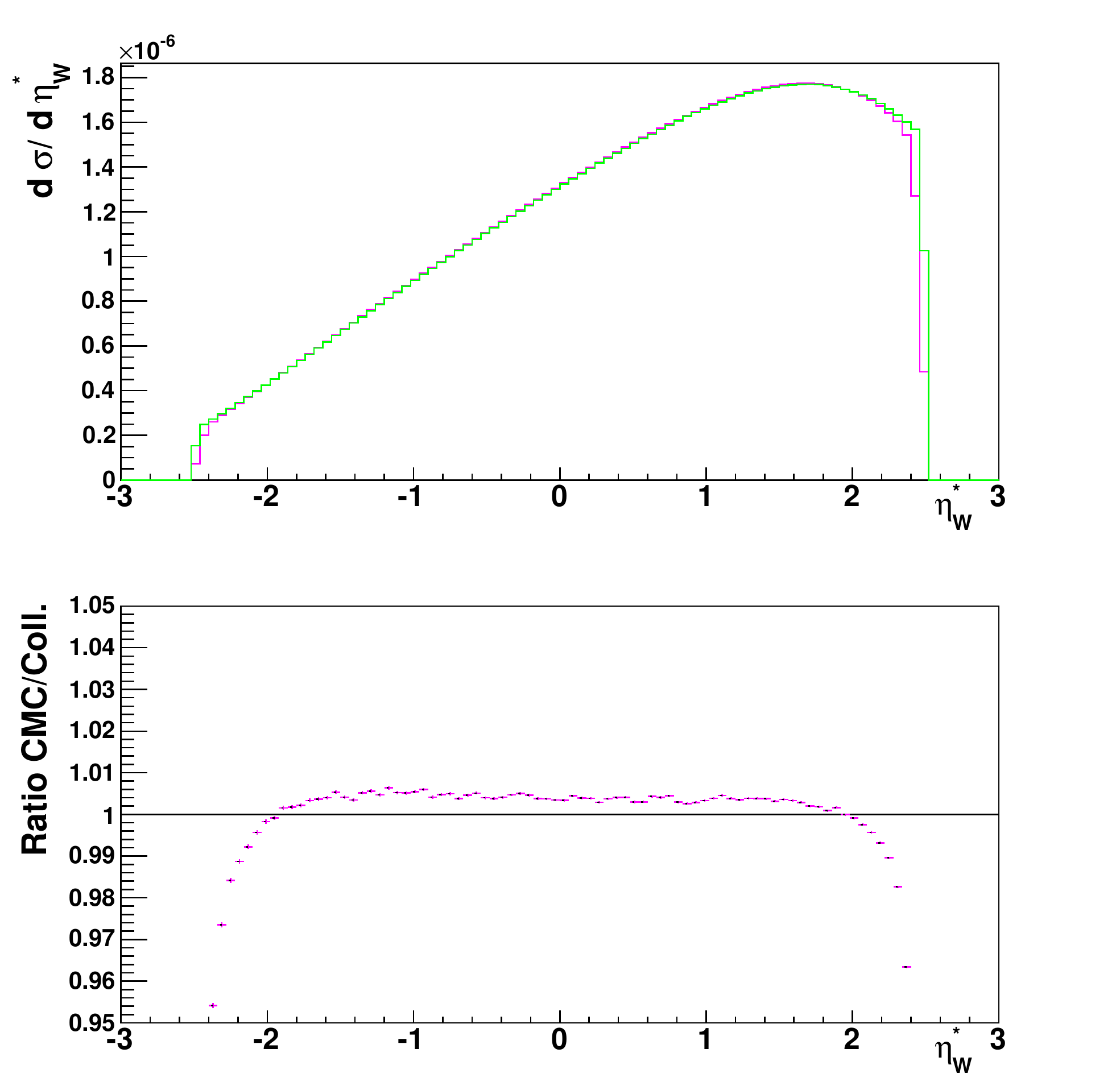}}
  \caption{
    In the upper plot
    the LO distribution of $\eta_W^*=\frac{1}{2}\ln(x_F/x_B)$ from the CMC
    LO parton shower (purple) and from the strictly collinear formula (green)
    are shown. The lower plot shows the ratio of the two,
    the agreement of $<0.5\%$ is obtained.
    }
  \label{fig:etaW_LO_7TeV}
\end{figure}
Figure~\ref{fig:etaW_LO_7TeV} 
represents a ``calibration benchmark'' for the overall normalization
at the LO level.
In fact, we show in Fig.~\ref{fig:etaW_LO_7TeV}
the properly normalized distribution of the variable
$\eta_W^*=\frac{1}{2}\ln(x_F/x_B)$.
In the collinear limit this variable represents the rapidity of $W$ boson.
This variable will differ substantially from the true rapidity of 
the $W$ boson in the presence
of the gluon with hard transverse momentum,
but this is an acceptable approximation in the current LO exercise.
The $W$ mass distribution with the sharp Breit-Wigner resonance lineshape
is not very interesting and we do not show it here.

In Fig.~\ref{fig:etaW_LO_7TeV} one of the distributions is from the MC
generation of the variables $(x_F,x_B)$ according to 2-dimensional
integrand of eq.~(\ref{eq:LOfactDY2LO}).
This is done using the general purpose MC program FOAM~\cite{foam:2002}.
However, in this MC we need the collinear PDF $D(t,x)$
in the entire range of $x$ and $t$ as an input.
This distribution has been obtained from a separate high statistics run
($10^{10}$ events) of a simple Markovian MC (MMC),
recording the resulting $D(t,x)$ in the 2-dimensional table (a finite grid).
In fact this MMC run solves the LO DGLAP equation 
(for gluonstrahlung LO kernel) using the MC method,
similarly as in refs.~\cite{Jadach:2008nu,GolecBiernat:2006xw}.%
\footnote{
 The use of the MC method is not mandatory here
 -- we could solve it using finite step methods,
 as in ref.~\cite{Botje:2010ay}.}
From the look-up table recorded during the MMC run,
a simple interpolation is employed to obtain $D(t,x)$ for any values 
of $t$ and $x$ in the next step,
that is in the 2-dimensional integrand used by FOAM.

Another distribution in Fig.~\ref{fig:etaW_LO_7TeV} comes
from the full scale MC generation 
(four-momenta conserving)
according to eq.~(\ref{eq:LOMCFBmaster}).
The MC run with $10^8$ events was used.
In the MC implementation we cannot use the Markovian method because
of the narrow Breit-Wigner peak due to a heavy boson propagator.
We could employ a backward evolution algorithm of 
ref.~\cite{Sjostrand:1985xi}, 
but instead we have opted to employ a variant of
the constrained MC (CMC) technique of ref.~\cite{Jadach:2007qa}.
In fact, we combine two CMC modules and FOAM into one MC generating
gluon emission from the incoming quark and antiquark which
annihilate into the $W$ boson.
The LO hard process ME of the $W$ boson production is implemented,%
\footnote{In the presend MC exercise the average over angular distribution
   of the $W$ boson decay products is taken. This is irrelevant
   for the conclusions of our study and this averaging can be undone
   rather easily.}
FOAM is taking care of the generation of the variables
$x_F,x_B,x_{F0},x_{B0}$ and the sharp Breit-Wigner peak
in $\hat{s}=s x_F x_B$, then initial paremeters for two CMC
modules are set and the gluon four-momenta $\bar{k}^\mu_j$ are generated.
Once they are mapped into $k^\mu_j$, following the prescription
defined in ref.~\cite{Jadach:2011cr},
the overall energy-momentum conservation is achieved.

Figure~\ref{fig:etaW_LO_7TeV} demonstrates
a very good numerical agreement
between $d\sigma/ d \eta_W^*$ 
from our full scale LO parton shower MC
of eq.~(\ref{eq:LOMCFBmaster})
and the simple formula of eq.~(\ref{eq:LOfactDY2LO})
in the strict collinear kinematics
(just convolution of two PDFs and the Born cross section).
The LO MC is working
in the standard phase space, with the exact 4-momentum conservation
and agrees with precision $<0.5\%$ with the simple collinear formula 
of eq.~(\ref{eq:LOfactDY2LO}).
The visible numerical bias is most likely due to finite size
of the grid used to parametrize PDFs from the MMC run.

\subsection{Numerical test of NLO correction}
Having cross-checked very precisely the overall normalization of our LO MC,
we are now ready to do a similar cross-check 
in case of the NLO-corrected hard process.

\begin{figure}[!t]
  \centering
  {\includegraphics[width=0.75\textwidth,height=85mm]{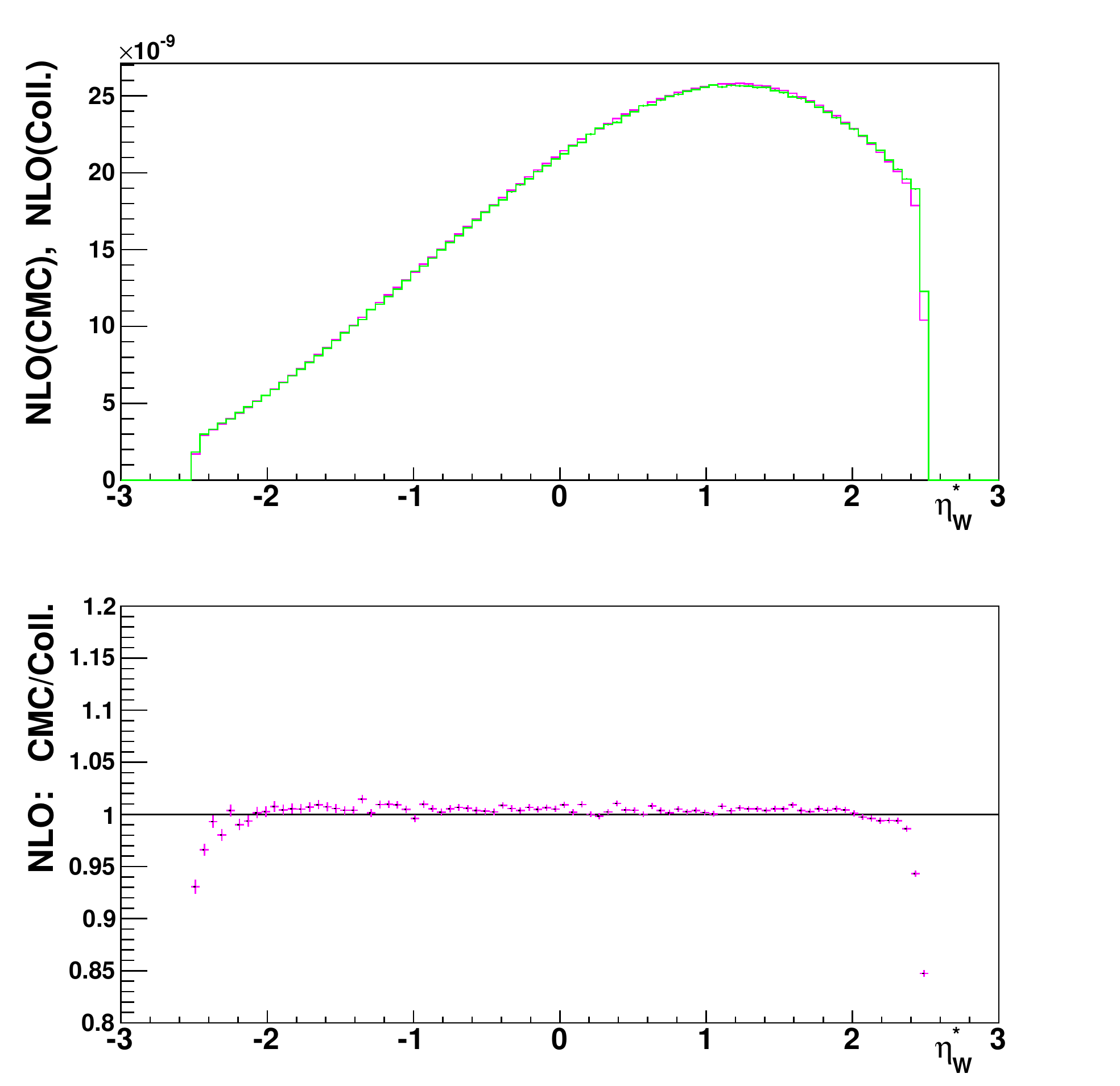}}
  \caption{
    The pure $(-)$~NLO correction to the distribution of 
    $\eta_W^*=\frac{1}{2}\ln(x_F/x_B)$ in CMC
    LO parton shower in $W$ boson production (purple).
    It agrees with the strictly collinear formula (green) to within $<1\%$
    of the NLO correction itself.
    }
  \label{fig:etaW_NLOglu_7TeV}
\end{figure}

\begin{figure}[!t]
  \centering
  {\includegraphics[width=0.75\textwidth,height=85mm]{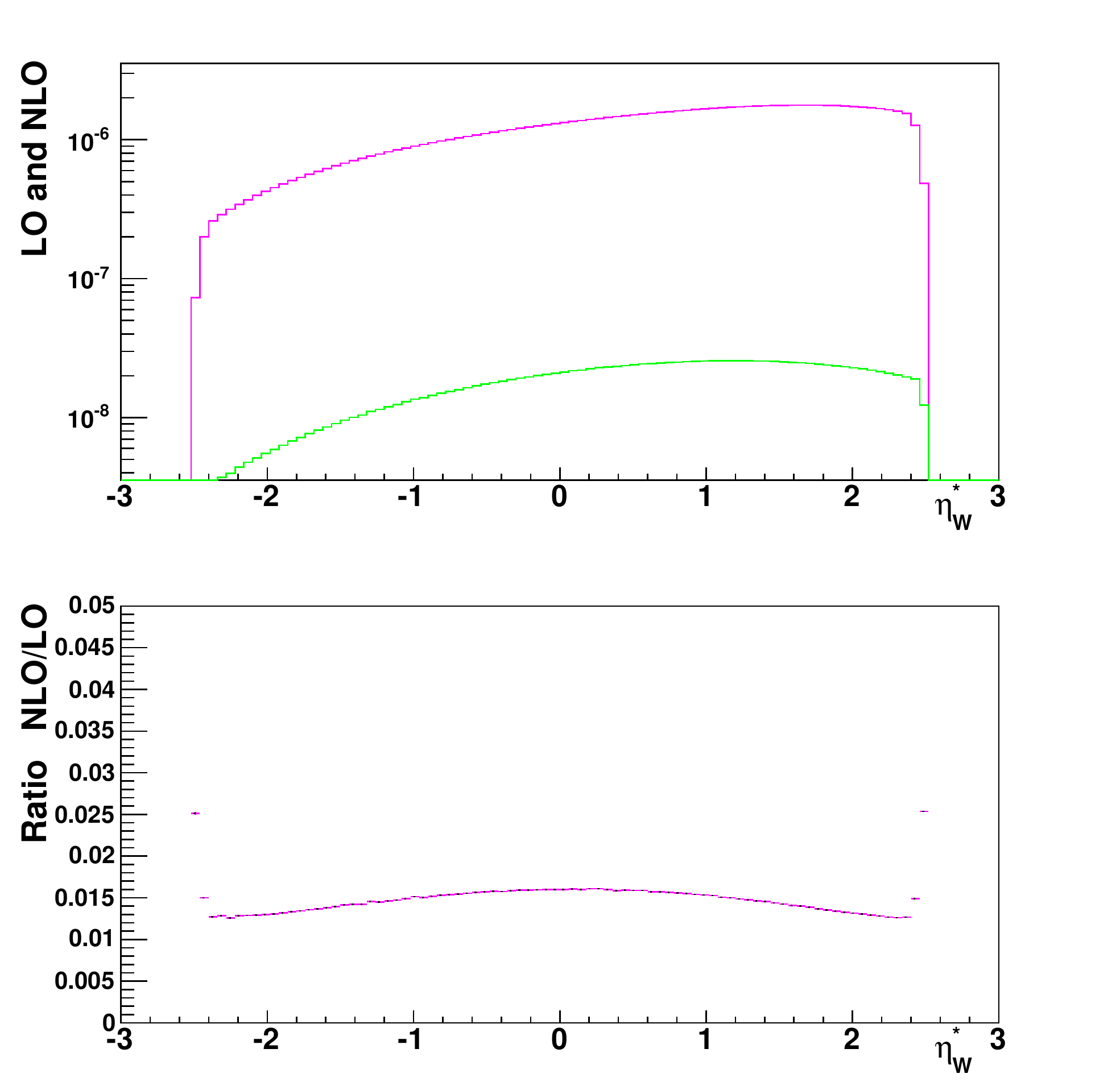}}
  \caption{
    The comparison of the LO (purple) and of
    the pure $(-)$~NLO corrections (green) to 
    the distribution of $\eta_W^*=\frac{1}{2}\ln(x_F/x_B)$;
    the overall normalization is in GeV$^{-2}$.
    }
  \label{fig:etaW_CanV7n_7TeV}
\end{figure}

Figure~\ref{fig:etaW_NLOglu_7TeV} represents a principal 
(technical) test and {\em proof of concept} of our new
methodology for implementing the NLO corrections to the hard process
in the parton shower MC.
The NLO correction to the $\eta_W^*$ distribution is obtained
on one hand within the full scale parton shower MC featuring
the NLO-corrected hard process as
in eqs.~(\ref{eq:LOMCFBmaster}) and~(\ref{eq:NLODYMCwt}),
and, on another hand,
with a simple collinear formula of eq.~(\ref{eq:DYanxch})
in which two PDFs are convoluted with
the analytical function $C_{2r}(z)$,
the ``coefficient function'' for the hard process.
In Fig.~\ref{fig:etaW_NLOglu_7TeV} we present the NLO corrections
obtained using both calculations.
The LO component, cross-checked in the previous section,
is present in the MC but not shown in this plot
in order to increase the ``resolution''.
In Fig.~\ref{fig:etaW_NLOglu_7TeV} we also include
the ratio of the NLO corrections from the two sources.

As seen in Fig.~\ref{fig:etaW_NLOglu_7TeV},
the result of the parton shower MC 
with the NLO-corrected hard process and the result
of the simple collinear formula of eq.~(\ref{eq:DYanxch})
agree very well, within the statistical error.

In Fig.~\ref{fig:etaW_NLOglu_7TeV} we
see only the NLO corrections, but
how big is the NLO correction with respect to the LO?
We show this in Fig.~\ref{fig:etaW_CanV7n_7TeV}, where
both the LO and NLO components are compared, and the NLO/LO ratio
is plotted as well.
As we see, the NLO correction to the rapidity-like variable
for the $W$ boson
is only about 1.5\% of LO, and this is unusually small.

For this particular hard process the NLO correction is negative,
hence in both Figs.~\ref{fig:etaW_NLOglu_7TeV} and \ref{fig:etaW_CanV7n_7TeV}
it is multiplied by the factor $(-1)$, in order to facilitate
visualization of the results.

\begin{figure}[!t]
  \centering
  {\includegraphics[width=0.75\textwidth,height=55mm]{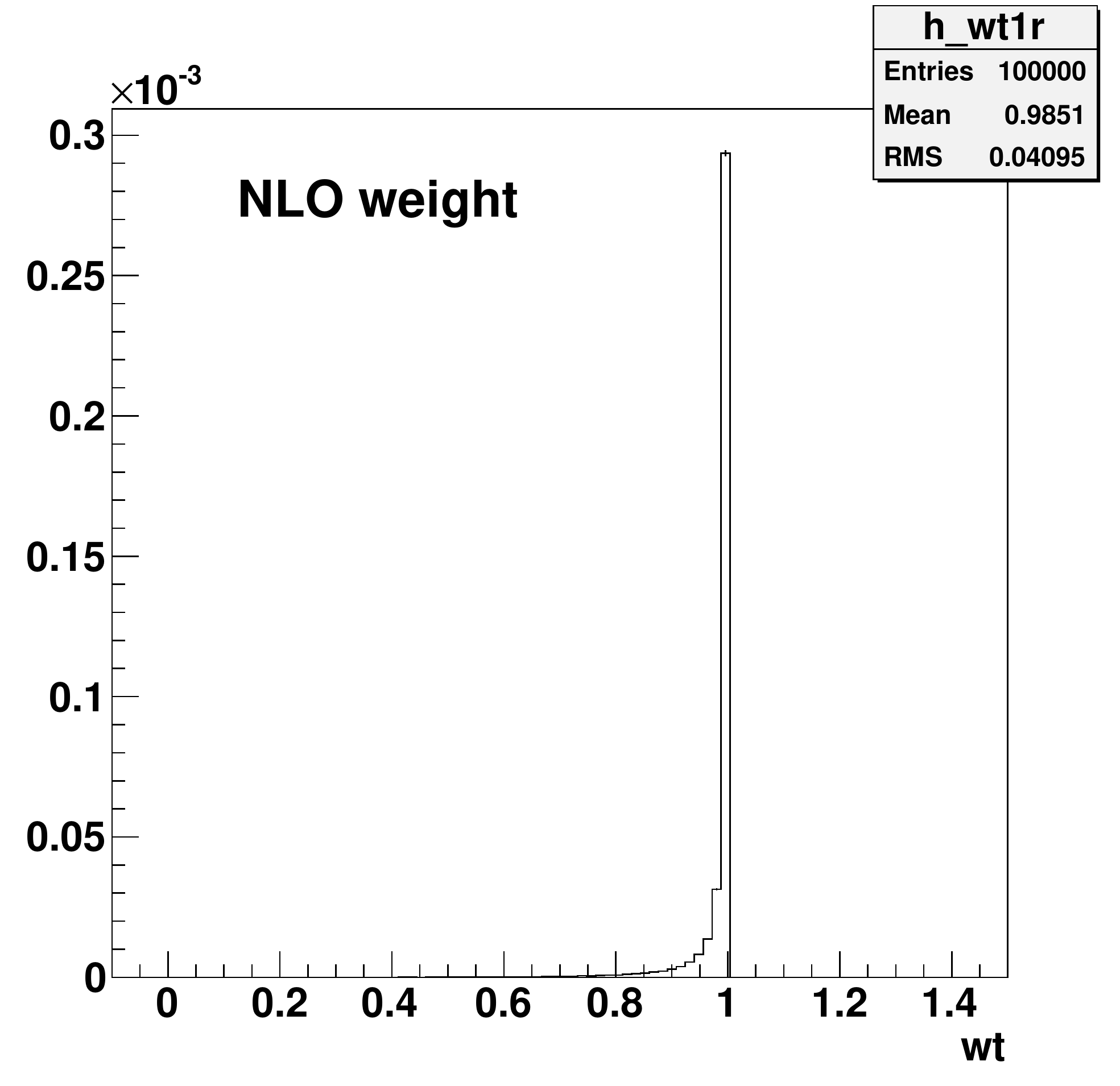}}
  \caption{
    The distribution of the NLO weight 
    $W^{NLO}_{MC}$ of eq.~(\ref{eq:NLODYMCwt}).
    }
  \label{fig:Canv1w_7TeV}
\end{figure}

At the technical level, the inclusion of the NLO correction
in the parton shower MC is straightforward, we are just activating
$W^{NLO}_{MC}$ of eq.~(\ref{eq:NLODYMCwt}).
In fact MC is providing the LO and NLO-corrected results in a single MC run
with weighted events.
The $W^{NLO}_{MC}$ weight is well behaved, strongly peaked near
$W^{NLO}_{MC}=1$, positive, and without long-range tails.
The distribution of this
weight is shown in Fig.~\ref{fig:Canv1w_7TeV}.
In the MC implementing the collinear formula of eq.~(\ref{eq:DYanxch}),
we again use FOAM, but now the generation space is 3-dimensional
due to the presence of additional variable $z$.

Note that in all numerical results shown so far we have put
$\Delta_{V+S}=0$, as it is completely unimportant for
the purpose of the presented numerical analysis.
For our study we also used toy initial distributions
$d_0(q_0,t)$, which were parametrize as follows:
\[
\begin{split}
xD_{q\in p}(x,1\text{GeV}) &= 2xu(x)+xd(x)+\frac{1}{2}xs(x),
\\
  2xu(x) &= 2.19\; x^{1/2} (1-x)^3,\quad
\\
   xd(x) &= 1.23\; x^{1/2} (1-x)^4,
\\
   xs(x) &= 1.35\; x^{0.2} (1-x)^7.
\end{split}
\]

\section{Discussion and comparison with other methods}

The new method of introducing the NLO corrections in
the hard process proposed in ref.~\cite{Jadach:2011cr}
and tested in this work is clearly very different
from the well established
MC@NLO  \cite{Frixione:2002ik}
and POWHEG \cite{Nason:2004rx,Frixione:2007vw}
methodologies.
Ref.~\cite{Jadach:2011cr} offers a limited
discussion on these differences.
Having at hand MC numerical implementation we may elaborate
on certain issues in more detail,
in particular we are going to show numerical
results illustrating differences with the POWHEG technique.

At first sight, the most striking difference 
with the POWHEG and MC@NLO techniques are:
\begin{itemize}
\item
``Democratic'' summation over all emitted gluons,
without deciding explicitly
which gluon is the one involved in the NLO correction
and which ones are merely ``LO spectators'' in the parton shower.
\item
The absence of $(1/(1-z))_+$ distributions 
in the real part of the NLO corrections 
(kinematics independence of the virtual+soft correction).
\end{itemize}

In the following, we shall elaborate mainly on the first point,
analyzing in a detail how 
$W^{NLO}_{MC}$ of eq.~(\ref{eq:NLODYMCwt})
is distributed over the multigluon phase space.
In order to make the discussion maximally transparent,
let us consider a simplified weight
\begin{equation}
\label{eq:NLODYMCwt_sim}
\begin{split}
&W^{NLO}_{MC}= 1+\sum_{j\in F} W^{NLO}_j,\qquad
W^{NLO}_j=
 \frac{\tbet_1(q_1,q_2,\bar{k}_j)}%
      {\bar{P}(z_{Fj})\;d\sigma_B(\hat{s},\hat\theta)/d\Omega}
\end{split}
\end{equation}
which is limited to one ladder (one hemisphere).
Moreover we put it on top of the
MC modeling gluon emissions from single quark,%
\footnote{We use the Markovian MC implementation,
  but optional use of the CMC would provide identical results.
  We use quite a wide range of $t$,
  corresponding to $\sqrt{s}=7$TeV.}
essentially the multigluon distribution of eq.~(\ref{eq:LOMC}).

\begin{figure}
  \begin{centering}
  \subfloat[]{
  \includegraphics[width=0.48\textwidth]{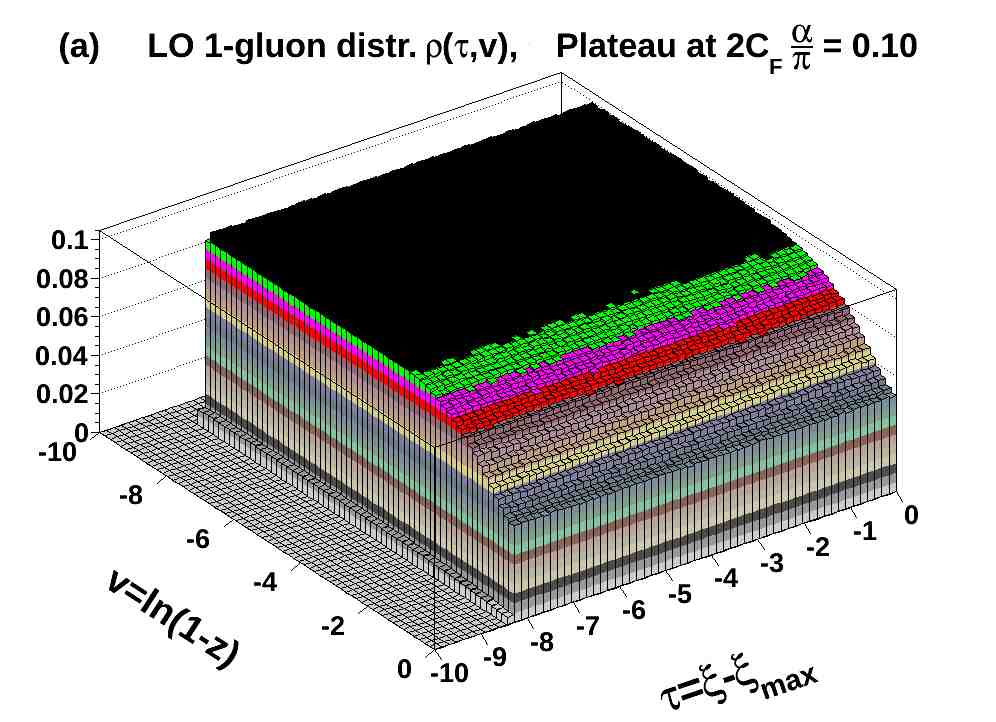}
  \label{fig:mcRho1gluLO_LO}
  }
  \subfloat[]{
  \includegraphics[width=0.48\textwidth]{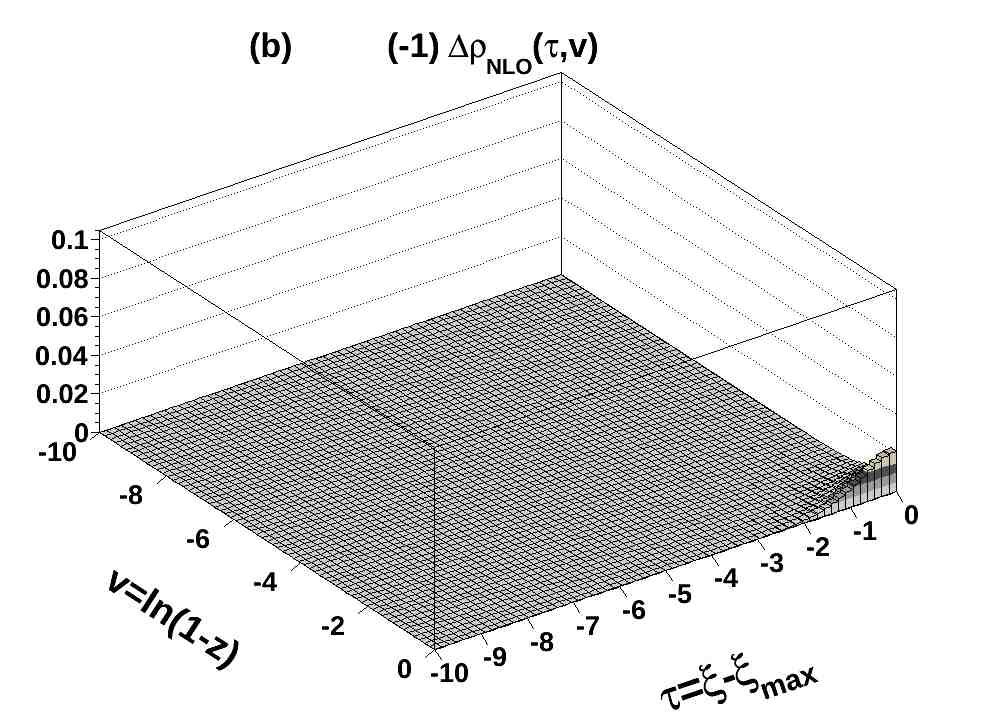}
  \label{fig:mcRho1gluLO_NLO}
  }
  \caption{
    (a) The inclusive distribution of gluons on the log Sudakov plane
    of rapidity $t=\xi_{\max}$ and $v=\ln(1-z)$.
    (b)~Contributions from all gluons weighted
    with the component weight $W^{NLO}_j$.
    }
  \label{fig:mcRho1gluLO}
  \end{centering}
\end{figure}

We start by examining the inclusive distribution of gluons
on the Sudakov logarithmic plane of rapidity $\xi$
and variable $v=\ln(1-z)$.
This is shown in Fig.~\ref{fig:mcRho1gluLO_LO}.
The distribution looks as expected,
and the flat plateau represents IR singularity 
$2C_F\frac{\alpha_S}{\pi} d\xi\frac{dz}{1-z}$
(for constant $\alpha_S$)
with the drop by factor 1/2 towards $z=0$,
due to $\frac{1+z^2}{2}$ factor in the LO kernel.

In Fig.~\ref{fig:mcRho1gluLO_NLO}
we show contributions from all gluons weighted
with the component weight $-W^{NLO}_j$
of eq.~(\ref{eq:NLODYMCwt_sim}).
(We insert a minus sign in order to facilitate visualization.)
Here we see that the NLO contribution is concentrated in
the area near the rapidity of the hard process $t=\xi_{\max}$,
which has to be true for the genuine NLO contribution.
On the other hand, the fact that the NLO correction dies out
towards the IR limit $z\to 1$ is not guaranteed in the collinear
factorization.
It results from the conscious choice that our LO differential distributions
reproduce the correct IR limit not only in LO but also in NLO
and in the entire phase space.%
\footnote{Older version of the standard LO MCs do not always
   reproduce the correct soft gluon limit beyond the LO level.}

Another important point
is the completeness of the phase space near the
($z=0$, $t=\xi_{\max}$) phase space corner.
Both POWHEG and MC@NLO use standard LO MCs which
feature an empty ``dead zone'' in this region,%
\footnote{This is due to the use of the boost transformation
 in the standard LO MCs to get the overall four-momentum conservation.
 We avoid this transformation (problem) as we use
 the rescaling transformation only in $\bar{k_j}\to k_j$ in the LO MC.
}
which is critical for the completeness of the NLO corrections
in the hard process.
They have to fill in this empty part of the phase space
with MC events according to the correct LO+NLO distribution.
Correcting for this deficiency of the standard LO MC 
requires non-trivial effort.
In our case, the problem of the phase space incompleteness
is absent%
\footnote{Of course, reconstructing the LO parton shower also requires
  non-trivial effort.}
and we simply re-weight the LO distribution (MC events) to the NLO level.

Finally, we also see that the NLO correction is very small,
which might be a general feature of the new method.
It is mainly due to the absence of
the $(1/(1-z))_+$ terms in the NLO correction --
this is a separate issue discussed in ref.~\cite{Jadach:2011cr},
see also a few remarks below.

\begin{figure*}[!ht]
  \centering
  {\includegraphics[width=1.0\textwidth]{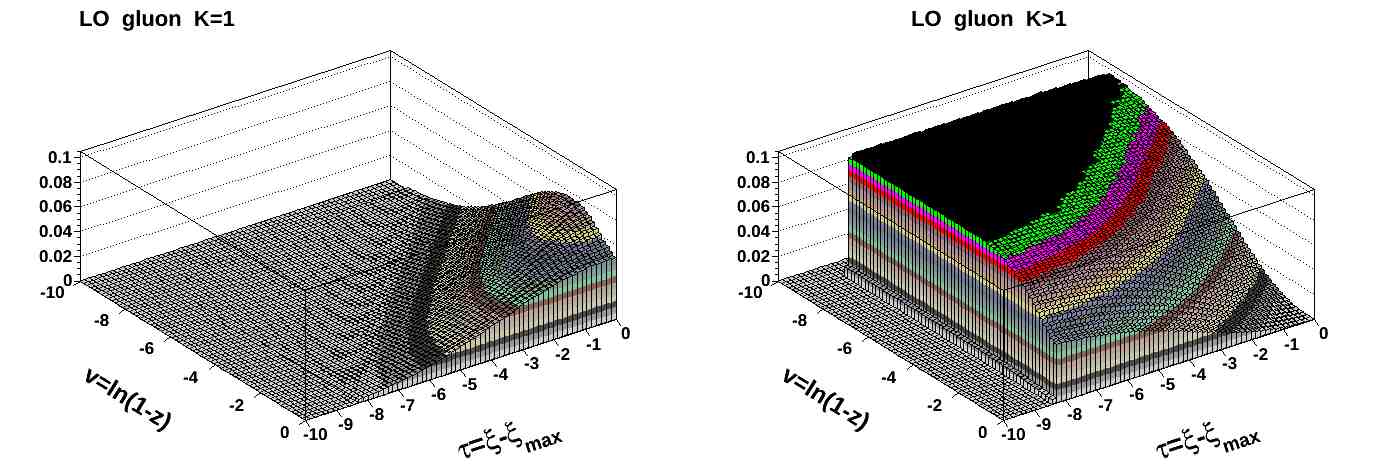}}
  \caption{
    The inclusive distribution of gluons of Fig.~\ref{fig:mcRho1gluLO_LO}
    split into the hardest (in $k^T$) gluon (left) and the rest (right).
    }
  \label{fig:mcCanv1k}
\end{figure*}

Looking at Fig.~\ref{fig:mcRho1gluLO} it is tempting to conclude that
the dominant contribution to $\sum_j W^{NLO}_j$ may come
from the gluon with the highest $\ln k_j^T\sim \xi_j+\ln(1-z_j)$,
that is the closest to the hard process corner ($z=0$, $t=\xi_{\max}$).
We may easily relabel gluons generated in the MC,
$\sum_j \to \sum_K$,
such that they are ordered in the variable $\kappa_K=\xi_K+\ln(1-z_K)$,
with $K=1$ being the hardest one ($\kappa_{K+1}<\kappa_K$).

In Fig.~\ref{fig:mcCanv1k} we show a split of the inclusive
distribution of Fig.~\ref{fig:mcRho1gluLO_LO}
into the $K=1$ component (hardest gluon in $k^T$) and the rest $K>1$.
As we see, the $K=1$ component saturates/reproduces
the original complete distribution
of Fig.~\ref{fig:mcRho1gluLO_LO} over all the region
where the NLO correction (Fig.~\ref{fig:mcRho1gluLO_NLO})
is non-negligible.
This is exactly the observation on which POWHEG technique is built!
Moreover, as noticed by the POWHEG authors, taking the $K=1$ component is
sufficient to reproduce the complete NLO correction
(up to NNLO).

\begin{figure}[!ht]
  \centering
  {\includegraphics[width=0.75\textwidth,height=55mm]{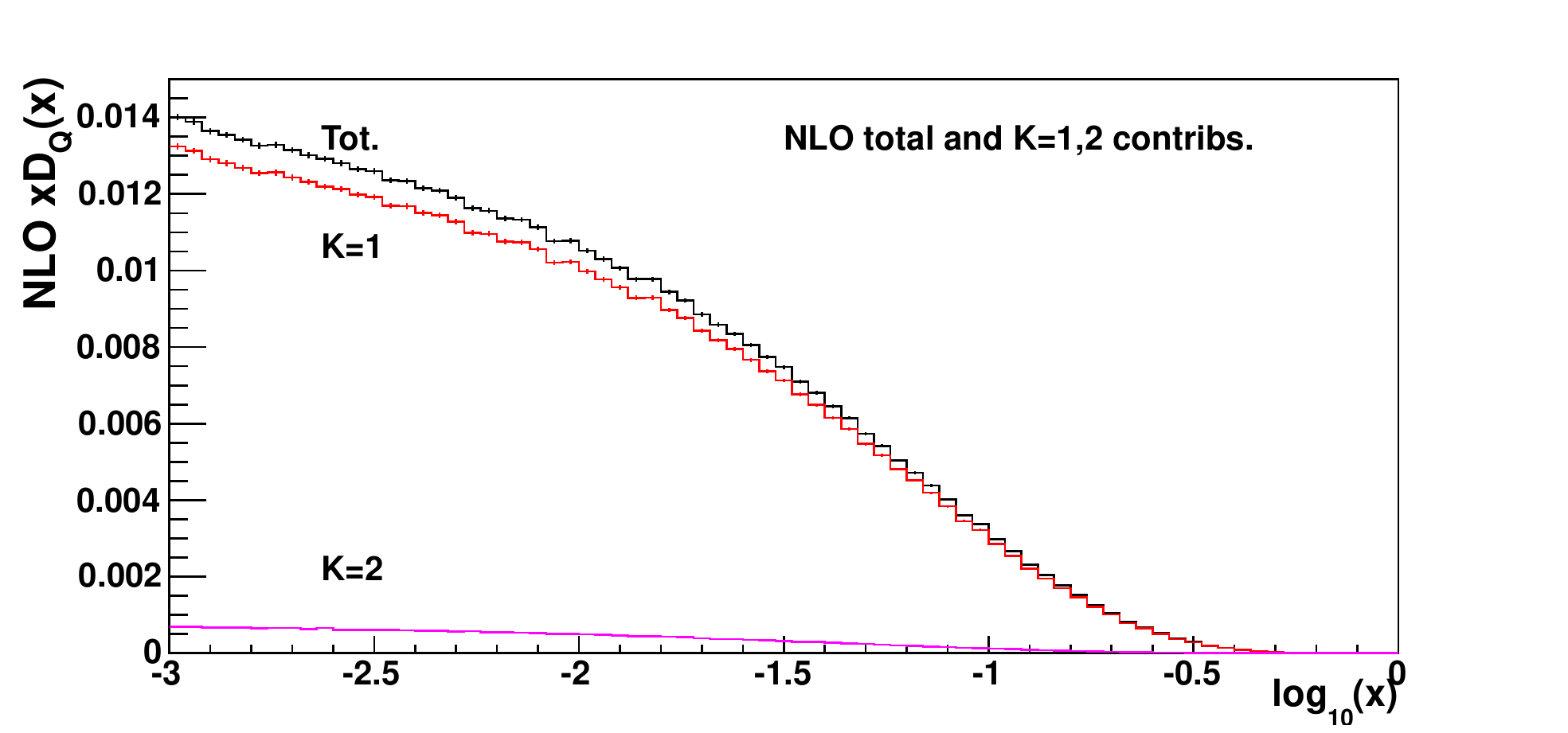}}
  \caption{
    The full NLO correction $\sum_j W^{NLO}_j$ 
    and its two hardest (in $k^T$)
    components $W^{NLO}_{K=1}$, $W^{NLO}_{K=2}$
    as a function of $x=\prod_j z_j$.
    }
  \label{fig:mcCanV8k}
\end{figure}

Let us check numerically the above statements 
by means of comparing the NLO correction
to the $x=x_0\prod_j z_j$ distribution from $\sum_j W^{NLO}_j$
and from $ W^{NLO}_{K=1}$.
This comparison is shown in Fig.~\ref{fig:mcCanV8k}.
As we see the $K=1$ component saturates the entire sum very well,
whereas the $K=2$ component is quite small.
A natural question is: why bother to keep the entire sum
instead of taking only the $K=1$ contribution?
In fact we can, which is valuable feature of our scheme.
However, we stress that
in the POWHEG scheme the $K=1$ gluon is generated in the MC
separately in the first step
and other gluons are generated (by the LO parton shower MC)
in the next step.
This is fine and easy if the LO MC uses $k^T$-ordering, while in case
of the LO MC with angular/rapidity-ordering additional effort
of generating the so called vetoed showers and truncated showers
is needed in the POWHEG method.
In our method, the angular ordering is used
but the vetoed/truncated showers are not needed,
even if we replace the sum $\sum_j W^{NLO}_j$ by $K=1$ 
component $W^{NLO}_{K=1}$.
Is there any rationale for keeping the sum over gluons in NLO weight at all?
There are two reasons for keeping it, at least optionally:
(a) the valuable crosscheck of the NLO MC against the simple collinear
formula of eq.~(\ref{eq:DYanxch}) is exact only if we keep the sum,
(b) it may turn out that keeping the sum reduces missing NNLO corrections.
In our opinion one should keep both versions and check which one
better fits the complete NNLO or better agrees with additional
resummations beyond LO.

\begin{figure*}[!ht]
  \centering
  {\includegraphics[width=1.0\textwidth,height=90mm]{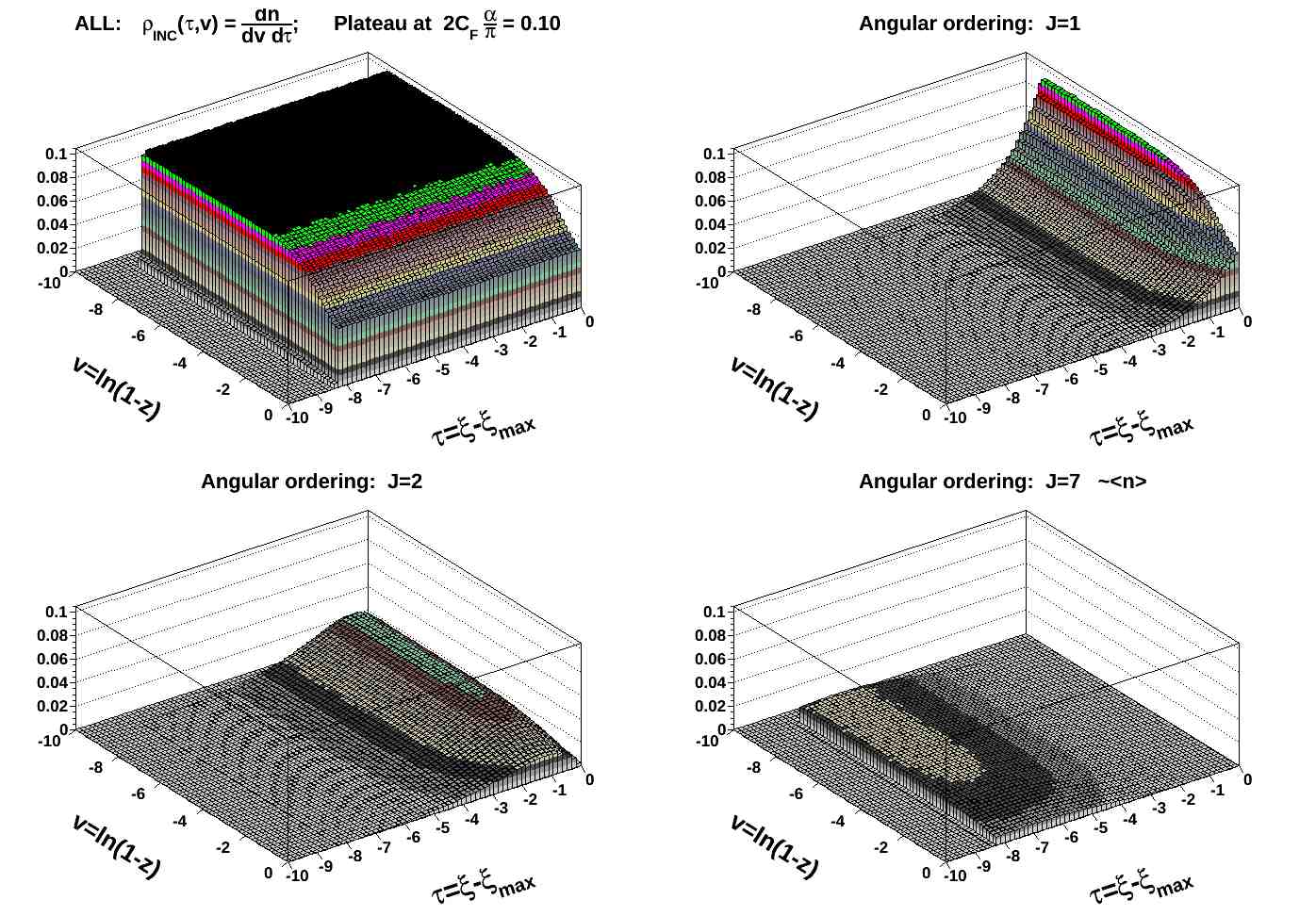}}
  \caption{
    The distribution of gluons ordered in rapidity,
    as in our basic LO MC.
    }
  \label{fig:mcRhoOrdLO}
\end{figure*}

As an additional illustration for the above discussion,
in Fig.~\ref{fig:mcRhoOrdLO} we show 
the distribution of gluons ordered in rapidity,
starting from the gluon with the maximum rapidity, 
the closest to hard process.
As we see the gluon distribution
with the highest rapidity $\xi\sim \xi_{\max}$ ($J=1$)
features a ridge extending towards the soft region.
It is important to notice that the width of this ridge
goes to zero when $\epsilon\to0$ in the IR cut-off $(1-z)<\epsilon$.
Hence, sooner or later the gluon with the highest $\xi$ will not be able
to reproduce/saturate the gluon distribution in the NLO corner
close to hard process, and the NLO correction will be highly incomplete.
In the $k^T$-ordering this is not the case, of course.

In the above, we have mainly discussed the differences
of our MC with the POWHEG
method in which, similarly to our case,
the negative MC weight is not allowed.
The MC@NLO method roughly corresponds to generating in a separate
MC branch events according to non-positive NLO correcting distribution.
In the same MC branch, events filling the empty phase space near
the hard process corner are also added.%
\footnote{This luckily reduces the number of events with
the negative weight.}

Both MC@NLO and POWHEG feature the $(1/(1-z))_+$ components in the
NLO corrections, which are the source of the practical complications
there, while they are absent in our approach.
In MC@NLO and POWHEG case these $(1/(1-z))_+$  corrections
act effectively as the ``in-flight'' translation of the PDFs
from the $\overline{MS}$ collinear factorization scheme (FS)
to the FS used effectively in the MC,
(see ref.~\cite{Jadach:2011cr}).
We propose to shift this translation beyond the MC,
as a rather simple redefinition of PDFs which should be done
``off line'', from the point of view of MC.
The above issue requires a dedicated study (in preparation),
and is also closely related to the upgrade of the ladder part
of the MC to the NLO level.

Having in mind that the considered method is more general,
and can be also applied for introducing the NLO corrections
in the middle of the ladder~\cite{Jadach:2010ew,Skrzypek:2011zw},
it is an interesting question
whether limiting the sum $\sum_j W^{NLO}_j$ to one (or two)
terms would/could be used in order to upgrade the QCD
evolution in the parton shower to the complete NLO level,
which would open many new promising avenues in the development
of the high quality QCD parton shower MCs for LHC and other colliders.
This question will be
addressed in the forthcoming study in ref.~\cite{IFJPAN-IV-2012-7}.

The present work provides a numerical crosscheck
of the ideas outlined in ref.~\cite{Jadach:2011cr}.
We briefly mention the most urgent future studies
which will necessarily follow this work 
(some are already completed but unpublished).
The two most important issues are:
(i) adjusting the choice of $\Xi$ at NLO level, and
(ii) selecting a better choice/definition of initial PDF.
Also, adding missing graphs for the NLO corrections,
that is graphs with gluon to quark transitions is needed.
This should be simpler than the presented gluonstrahlung
contributions due to lack of IR singularities.

The present choice of the rapidity boundary $\Xi=0$
is good at the LO level, and it also correctly reproduces the integrated NLO
cross sention. 
However, at the exclusive level, 
the exact NLO distributions
must be properly reproduced in the limit
when all gluons but one are collinear (have small $k^T)$,
for instance, the rapidity difference between the heavy boson and the hardest
(in $k^T$) gluon.
For the above aim the best choice is to identify $\Xi$ with the rest
system of the heavy boson and the hardest gluon $\eta^*$.
This can be easily obtained by means of refining the mapping 
$\bar{k}_i\to k_i$ in such a way that it is used twice.
For the first time with $\Xi=0$, then the rapidity $\eta^*$ is determined
and $\bar{k}_i\to k_i$ mapping is repeated with $\Xi=\eta^*$.
Obviously some gluons will be reclassified as belonging to another
initial beam ladder.%
\footnote{Luckily, this ``flow'' of gluons from one to another hemisphere does
  not influence the overall MC weight.}
The above solution was already tested and works correctly.

Concerning further refinements on the initial PDF,
this issue would be resolved automatically if MC was fitted
to the experimental data
or if the PDFs have been fitted within the MC scheme.
If the initial PDF is to be taken from a standard library of
PDFs in the $\overline{MS}$ scheme, 
then it will be necessary to correct it
using the difference of the counterterms of the $\overline{MS}$ and MC schemes
(see eq.~(44) in \cite{Jadach:2011cr}).
From the classic analysis~\cite{Altarelli:1979ub}
of NLO corrections to the DY process, it is known that
this correction will be large and dominated by the term
$\sim \Big( \frac{\ln((1-z)^2/z)}{1-z} \Big)_+$,
in the region where quark distribution%
\footnote{Similar phenomenon will occurs for the gluon distribution.}
is strongly varying in $x$.
Note that in POWHEG method the above correction is implemented in the MC
by means of explicit generation of the variables in the convolution
implementing NLO corrections and the corresponding manipulation
on four-momenta is done.
In contrast, in our method NLO corrections are included entirely
through MC weight and no extra kinematics transformations are needed
(beyond these of the LO MC modelling).

In the above context an interesting numerical results are presented in
ref.~\cite{Hautmann:2012dw}
-- they illustrate size and location of the $x$-variation in PDFs
due to kinematics manipulations in POWHEG driven by NLO corrections.
In our method the entire kinematical modification of the longitudinal
parton fraction $x$ is due to the LO mapping and shape modification
due to the NLO weight.
We expect this effect to be less sizable in our method, but a separate
study would be needed to verify it.

\section{Summary and outlook}
A new methodology of adding 
the QCD NLO corrections to the hard process
in the initial state Monte Carlo parton shower
is tested numerically using
heavy boson production
at hadron-hadron colliders.
The ladder parts of the parton shower are modelled in the LO
approximation, also using these new methods.
The presented numerical results prove
that the basic concept of the new methodology
works correctly in the numerical environment of 
a Monte Carlo parton shower.
The differences with the well established methods
of MC@NLO and POWHEG are briefly discussed.
Also, possible refinements of the method are indicated.

Clearly
the ''proof of concept'' is successful, and more work is
required before practical application will emerge.



\section*{Acknowledgment}
This work is partly supported by 
the Polish Ministry of Science and Higher Education grant No.\ 1289/B/H03/2009/37,
 the Polish National Science Centre grant DEC-2011/03/B/ST2/02632,
 the NCBiR grant LIDER/02/35/L-2/10/NCBiR/2011,
  the Research Executive Agency (REA) of the European Union 
  Grant PITN-GA-2010-264564 (LHCPhenoNet),
the U.S.\ Department of Energy
under grant DE-FG02-04ER41299 and the Lightner-Sams Foundation.
One of the authors (S.J.) is grateful for the partial support
and warm hospitality of the TH Unit of the CERN PH Division,
while completing this work.



\end{document}